\documentclass[review]{elsarticle}

\usepackage[a4paper, margin=2.5cm]{geometry}

\journal{Safety Science}

\usepackage{lineno,hyperref}
\usepackage{numprint}
\npthousandsep{,}\npthousandthpartsep{}\npdecimalsign{.}
\modulolinenumbers[5]

\usepackage{color}
\usepackage{comment}
\usepackage{enumitem}
\usepackage{tabularx}
\usepackage{array}
\usepackage{color, colortbl}
\usepackage{multirow}
\usepackage{longtable}
\usepackage{pdflscape}
\usepackage{numprint}
\usepackage{amssymb}
\usepackage{caption}
\usepackage{amsmath}
\usepackage{arydshln}

\definecolor{Gray}{gray}{0.9}

\newcommand{\quotes}[1]{``{#1}''}
\renewcommand{\arraystretch}{0.8} 

\usepackage{fancyhdr}
\begin{document}

\newcolumntype{R}[1]{>{\raggedright\let\newline\\\arraybackslash\hspace{0pt}}m{#1}}
\newcolumntype{C}[1]{>{\centering\let\newline\\\arraybackslash\hspace{0pt}}m{#1}}
\newcolumntype{L}[1]{>{\raggedleft\let\newline\\\arraybackslash\hspace{0pt}}m{#1}}

\begin{frontmatter}
\title{The status quo of fire evacuation drills in nursery schools\footnote{This is a preprint version of the article accepted to Safety Science with DOI 10.1016/j.ssci.2025.106915}}

\pagestyle{fancy}

\author[fsv]{Hana Najmanov\'a}
\ead{hana.najmanova@cvut.cz}
\author[fit]{Petr Nov\'ak}
\ead{petr.novak@fit.cvut.cz}
\author[lund]{Enrico Ronchi}
\ead{enrico.ronchi@brand.lth.se}
\cortext[cor1]{Corresponding author. Tel. +420224357151, email: hana.najmanova@cvut.cz}
\address[fsv]{Czech Technical University in Prague, Faculty of Civil Engineering, Th\'akurova 7, 166 29 Prague, Czechia}
\address[fit]{Czech Technical University in Prague, Faculty of Information Technology, Th\'akurova 9, 166 29 Prague, Czechia}
\address[lund]{Lund University, Department of Fire Safety Engineering, John Ericssons v\"ag 1, 22363, Lund, Sweden}

\begin{abstract}
Fire drills are a commonly-used training method for improving how people act in emergency situations. This article deals with fire drills in early childhood education facilities and analyses data gathered from an online survey of \numprint{1151} Czech nursery schools (23.5\% of nursery schools invited to participate, 21.5\% of all officially registered nursery schools in Czechia). It provides recommendations for improving the effectiveness of fire drills in nursery schools. Results suggest that regular (typically annual) fire drills are common training practices at these schools, but more frequent fire drills led to significantly fewer issues during evacuation. The most frequent evacuation issue encountered during fire drills, according to our data, was slow movement. Many children needed assistance, notably on stairs; nursery schools located only on the ground floor reported fewer issues during evacuation than other schools. The purpose, design, and implementation of fire drills with our study confirming results from prior studies must be properly integrated into complex and systematic fire safety education programs that specifically reflect the characteristics, needs, and possible limitations of preschool children.
\end{abstract}


\end{frontmatter}
\thispagestyle{fancy}

\paragraph{Highlights}
\begin{itemize}
    \item Online survey results on fire drills with \numprint{1151} Czech nursery schools.
    \item Operational conditions and real experiences with fire drills in nursery schools probed. 
    \item Factors affecting fire drills analyzed. 
    \item Slow movement and assistance required by children reported as the most common issues.
    \item Suggestions for improving the effectiveness of fire drills in nursery schools.
\end{itemize}

\newpage

\section{Introduction}
\label{sec:intro}
Fire safety design is one of the many factors that ensure safe and high quality environments throughout a building's life cycle \cite{gehandler_theoretical_2017}. Fire drills (evacuation drills for improving fire incident readiness) are a vital component of fire safety prevention measures and are generally recommended as a useful tool for improving how occupants of a building reach safety in emergency situations \cite{gwynne_enhancing_2019}. While there are many different fire drill objectives (e.g., improving evacuation procedures, giving participants and observers useful information via safe simulated emergency situations), two key purposes are: 1) training building occupants and safety personnel for evacuation readiness and 2) evaluating safety by observing fire drill performance \cite{noauthor_2021_2020}. From a research perspective, fire drills have become a frequently used tool for collecting and studying data about human behavior, including evacuation performance in fires \cite{gwynne_improving_2013, lovreglio_pre-evacuation_2019}. \par
Current research indicates that the potential benefits of fire drills are not always fully exploited. Limitations of routine fire drills include non-representative conditions and/or evacuation scenarios, inconsistent practices, and a lack of systematic data collection \cite{gwynne_pros_2016, catovic_survey_2018, kinateder_where_2021}. Less-than-ideal fire drill performance can also be the result of high financial and/or organizational costs and safety and ethical constraints that might possibly be mitigated using other fire training and assessment methods (e.g., serious games, virtual and augmented reality simulations) \cite{menzemer_scoping_2023, lovreglio_need_2017}. Since fire drills are the most common and traditional method for evaluating safety procedures, researchers are seeking deeper insights and understanding fire drills validity \cite{amos_need_2019}. A challenging shift towards evidence-based approaches \cite{gwynne_future_2020} and standardized reporting for the design and evaluation of fire drills \cite{baig_empirical_2024} have begun to emerge in recent fire safety research. \par
Fire drills are usually required by building regulatory frameworks that vary internationally in terms of frequency and participation requirements. Variations depend mostly on types of buildings/occupants involved, potential fire hazards, and other aspects \cite{nfpa_nfpa_2020, noauthor_2021_2020, gwynne_enhancing_2019}. Preschool education buildings accommodate a very high proportion of young children and may have a high ratio of adults to children (the so-called adult/staff-to-child ratio).\footnote{For example, a staff-to-child ratio of 1:8 means that one staff member is responsible for supervising 8~children. Note that a staff-to-child ratio of 1:8 is considered a higher value than staff-to-child ratio of 1:4 in this study.} 
Preschool children, typically ranging from 3~to 6~years of age, are generally considered to be a vulnerable population with limited self-rescue capabilities \cite{berk_child_2006, payne_human_2012}. This results in evacuation processes that often depend on the decisions and actions taken by responsible adults \cite{kholshevnikov_study_2012, fang_experimental_2019, najmanova_experimental_2023-1}. Such challenging evacuation conditions are reflected in national building codes and fire safety standards in the form of special or additional safety requirements for preschools. Legislative requirements on the regular performance of fire drills in early childhood education institutions vary worldwide. An interesting overview of this is provided in a study by Page and Norman \cite{page_prevalence_2015} focused on fire safety in early childhood centers housed in multi-story buildings (Table~\ref{tab:freq}). According to Czech legislation, fire drills are not explicitly required in preschool education institutions (i.e., nursery schools). Mandatory fire drills can be performed according to the prescribed or recommended frequencies in fire safety documentation filed by a professionally competent fire prevention representative. Beyond the large diversity in legislative requirements worldwide, more detailed information or guidance for the design, execution, and reporting of fire drills is generally not available \cite{gwynne_enhancing_2019}, resulting in inconsistencies in actual practices. \par 

\begin{table}[hbt!]
    \caption{Required frequency of fire drills in early childhood education institutions, reproduced from \cite{page_prevalence_2015} supplemented by Czech requirements}
    \label{tab:freq}
    \centering
    \scriptsize
    \begin{tabular}{C{1.5cm}|C{1.5cm}|C{1.8cm}|C{2.8cm}|C{2.2cm}|C{2.7cm}}
    \textbf{1$\times$ per} & \textbf{4$\times$ per} & \textbf{1--2$\times$ per} & \textbf{1$\times$ per year} & \textbf{Only} & \textbf{Not stipulated}  \\
   \textbf{month} & \textbf{year} & \textbf{year} &  & \textbf{recommended} &   \\
        \hline
    US (S), Ireland & US (M) & UK & Finland, Western Australia, Norway, Belgium & Denmark & Singapore, Sweden, Czechia \\ 
    \hline
    \multicolumn{6}{l}{\footnotesize US=United States, UK=the United Kingdom} \\
    \multicolumn{6}{l}{\footnotesize S=requirements for single-story buildings, M=requirements for multi-story buildings} \\  
    \end{tabular}
\end{table}

Observations of fire evacuation drills in early childhood educational settings have been widely used as data collection method for enhancing our knowledge on evacuation behavior and dynamics of young children in both toddler age (typically up to 3~years of age, \cite{larusdottir_step_2011, latosinski_assessing_2020}) and preschool age (typically ranging between 3~to 6~years of age, \cite{larusdottir_evacuation_2012, kholshevnikov_pre-school_2009, cuesta_collection_2016, najmanova_experimental_2017, najmanova_experimental_2023}). However, research studies using surveys to map fire safety conditions and performing evacuation drills in this particular type of occupancies remain scarce. 
Murozaki and Ohnishi \cite{murozaki_study_1985} conducted research concentrating on fire safety and evacuation measures in Japanese preschools and day care facilities. This study included a questionnaire survey distributed among 158~institutions that provided care to children age up to five years old. The results of the questionnaire revealed several insights into fire safety related issues such as staff-to-child ratio, frequency of fire drills and behavior of children during fire drills.  
Furthermore, Taciuc and Dederichs \cite{taciuc_determining_2014} conducted an international survey among daycare teachers and experts in child development to study self-prevention capabilities during evacuation in toddlers and young children. The survey study partially dealt with various fire safety installations, staff-to-child ratios, and frequency of fire drills in day-care centers in seven different countries.  

Our study examines specific conditions for conducting fire drills in preschool settings with children typically aged between 3~to 6~years of age. It presents survey data about daily operations and experiences with fire drills and emergency evacuations in \numprint{1151} Czech nursery schools (23.5\% of nursery schools invited to participate, 21.5\% of all officially registered nursery schools in Czechia). The paper provides background about the general characteristics of Czech nursery school environments relevant for fire safety design (e.g., data related to operational conditions, occupancies, capacities, and building enclosures) and describes actual practices and experiences with fire drills (including drill frequencies, execution, reporting, and issues encountered). Based on results from this survey, we then provide recommendations for improving the effectiveness of fire drills in nursery schools anywhere according to our investigation of the Czech context. 

\section{Research methods and background}
The data presented in this study were collected using an anonymous online survey with a questionnaire instrument. This method was considered an appropriate tool for the purpose of this study and has been used in the evacuation and fire safety field (\cite{haghani_empirical_2020, chen_elementary_2018}), including to inform decision-making processes \cite{elzie_groups_2023, nilsson_integrating_2018}, as an extension of evacuation drills and experiment surveys \cite{hulse_fire_2022,wang_heterogeneous_2023}, and to validate virtual reality experiments \cite{lovreglio_exit_2022, smedberg_impact_2023}. We collected data about real-life operating conditions, experiences, and practices for fire drills performed in nursery schools in Czechia. This section describes survey design, data gathering and analysis procedures, and the limitations of this investigation. This section also includes a brief description and summary of early childhood education and care in Czechia to provide relevant context for understanding and interpreting the results of this study. 

\subsection{Survey design and development}
The survey was designed by the research team to be completed by the leaders of the nursery schools listed in the Ministry of Education, Youth and Sports of Czechia's Register of Education Institutions (in total \numprint{5364}~registered nursery schools to June, 2019). \numprint{4930}~nursery schools with email contact information available in the Register were contacted to participate in the online survey via a link sent by email. The invitation was successfully delivered to \numprint{4903} nursery school leaders (91.4\% from the registered nursery schools). A survey tool in Czech was made available online in Google Forms from the June to July,~2019.

Before the survey email was sent, a participatory pilot survey was conducted with five nursery school leaders to obtain feedback on the design, clarity, and time spent on the survey. The pilot findings were used mainly to remedy content issues, such as improving the language of questions. Pilot responses were not included in the main analysis.

The survey presented here was part of a research project focused on the evacuation of preschool-aged children in Czech nursery schools. The research project involved different types of data collection and was approved by the Czech Technical University (CTU) in Prague’s Committee for Ethics in Research of the CTU Scientific Council under reference number 0000-01/19/51902/EKČVUT. It should be noted that this survey was anonymous and did not collect sensitive and confidential data about participants. As such, the anonymous survey data were deemed not to require further ethical scrutiny on top of the ethically approved research project.
The survey design was also vetted with the General Directorate of the Fire Rescue Service of Czechia's Prevention Department and the Department of Instruction and Training Services in order to determine what data would be useful in supporting their theoretical and practical use. 

The survey had two main sections:

\begin{itemize}
 \setlength\itemsep{0.1mm}
    \item \textbf{Part~I: General information} aimed at acquiring background information relevant for fire safety design, including questions about construction and operational conditions (e.g., building characteristics, occupancy characteristics and capacities); Table~\ref{tab:questions_I}. 
    \item \textbf{Part~II: Fire drills} in actual practice, probing participants' experiences in performing fire drills at their nursery schools; Table~\ref{tab:questions_II}. 
\end{itemize}

Part~I was intended for all nursery schools leaders invited to participate. Part~II was available only to nursery schools that performed fire drills according to what they reported in Part~I (Table~\ref{tab:questions_I}). The survey was designed to mitigate survey fatigue for participants while still gathering enough relevant data. Each section consisted of 10~questions (20~questions in total) with: 3~close-ended questions and 17~multiple-choice questions (both single and multiple selection answers were possible) with an editable \quotes{Other} category for 12~questions \cite{clark_brymans_2021}. A total of 19~questions allowed respondents to include additional information through a \quotes{Comments} field (referred to as \quotes{comment} in this study). None of the questions were mandatory. The survey instrument is presented in Table~\ref{tab:questions_I} and Table~\ref{tab:questions_II} (translated from Czech into English; questions are labeled \#1--20 for further reference).

\begin{table} [hbt!]
    \centering
    \footnotesize
    \caption{Survey instrument, Part~I}
    \label{tab:questions_I}
    \begin{tabular}{C{0.3cm}|R{11.3cm}|C{1.1cm}|C{1.6cm}}
    \multicolumn{3}{l}{\textbf{Part~I}} \\
    \hline
    \textbf{\#} & \textbf{Questions} & \textbf{Type} & \textbf{Responses (MIS/EXL) } \\
    \hline
    1 &  What region of Czechia is your nursery school located in?  & MC-SS & 1151 (0/0) \\
    \cdashline{2-2}
    & \textit{(Choice of 14~regions)} &  & \\
    \hline
    2 & How many classes are in your nursery school? & CE & 1148 (2/1) \\
    \hline
    3 & How many children, on average, are enrolled in one class? & CE & 1146 (5/0)  \\
    \hline
    4 & How many teaching staff (teachers, teacher's assistants, educators), on average, are  & CE & 1116 (8/27) \\    
    & present in a class? & \\
    \hline
    5 & Are children divided into classes based on their ages? If yes, please specify the age   & MC-SS & 1144 (7/0)\\   
    & categories. & \\
    \cdashline{2-2}
    & \textit{(No, classes are heterogeneous \textbar Yes, classes are homogeneous \textbar Other}) & \\
    \hline  
    6 & On which floors are the spaces intended for children in the nursery school building(s)?  & MC-MS & 1151 (0/0) \\ 
    \cdashline{2-2}
    & \textit{(Ground floor \textbar First floor \textbar Second floor \textbar Other}) & \\
    \hline
    7 & Is an external escape staircase(s) available in the nursery school building(s)? 
    & MC-SS &  1143 (8/0)\\ 
        \cdashline{2-2}
    & \textit{(Yes \textbar No)} & \\
    \hline
    8  & Is fire safety documentation (e.g., a report) available at your nursery school? & MC-SS & 1144 (7/0) \\
    \cdashline{2-2}
    & \textit{(Yes \textbar No \textbar I don't know)} $\ast$ & \\
    \hline
    9 & Is the issue of personal safety included in your nursery school's educational program?  & MC-SS & 1144 (7/0) \\   
    \cdashline{2-2}
    & \textit{(Yes \textbar No \textbar I don't know)} $\ast$ & \\
    \hline
    10 & Are fire drills performed at your nursery school?  & MC-SS & 1151 (0/0) \\
    \cdashline{2-2}
    & \textit{(Yes, at least once in the last 5~years \textbar No)} & \\
    \hline
    \multicolumn{4}{l}{\# =question number}, MIS=missing response, EXL=invalidated and excluded response \\ 
    \multicolumn{4}{l}{CE=close-ended question, MC-SS=single-select multi-choice question, MC-MS=multi-select multi-choice question} \\
    \multicolumn{4}{l}{$\ast$ Question included at request of the General Directorate of the Fire Rescue Service of Czechia}  \\
    \hline
    \end{tabular}
\end{table}    

\begin{table}[hbt!]
    \centering
    \footnotesize
    \caption{Survey instrument, Part~II}
    \label{tab:questions_II}
    \begin{tabular}{C{0.3cm}|R{11.5cm}|C{1.1cm}|C{1.6cm}}
    \multicolumn{3}{l}{\textbf{Part~II}} \\
    \hline
    \textbf{\#} & \textbf{Questions} & \textbf{Type} & \textbf{Responses (MIS/EXL)} \\
    \hline
    11 &  How often, on average, are fire drills performed in your nursery school? & MC-SS & 872 (8/0) \\
    \cdashline{2-2}
    & \textit{(1x a year \textbar 2x per year \textbar Other)} & \\
    \hline
    12 & How are fire drills announced to staff members?   & MC-SS & 880 (0/0)  \\
    \cdashline{2-2}
    & \textit{(All staff members know the date and time \textbar All staff members only know the date \textbar  } & \\
    & \textit{Fire drills are announced without specifying date or time \textbar Fire drills are not}  & \\
    & \textit{announced to staff members in advance \textbar Other}) & \\
    \hline
    13 & Where do teaching staff (teachers, teacher's assistants, educators) get instructions on & MC-MS & 880 (0/0) \\
    &   how to behave and act during fire drills (e.g.,~about evacuation route choice, how   & \\
    &  children are organized)?  & \\
    \cdashline{2-2}
    & \textit{(Regular fire protection education provided by a fire prevention officer \textbar Internal}  & \\
    & \textit{education provided by nursery school management (for example, by the school's leader)} & \\
    & \textbar \textit{Teaching staff are not specially trained/they act according to their professional}  & \\
    & \textit{experience \textbar Other)} & \\
    \hline
    14 & Which warning signal is used to start evacuation from the nursery school? Please, also  & MC-SS & 876 (4/0) \\  
    & specify its wording/sound.  & \\
    \cdashline{2-2}
    & \textit{(Voice message (verbal) \textbar Manual sound signal (e.g., whistle, gong, pot banging) \textbar}  & \\
    & \textit{Combination of voice message and manual sound signal \textbar \quotes{Coded message or sound}}   & \\
    & \textit{(e.g., a particular phrase or melody with the meaning known only to staff members) \textbar } & \\
    & \textit{Activation of the fire alarm system (e.g., a smoke detector) \textbar Other)} & \\
    \hline
    15 & How is the warning signal to begin evacuating disseminated through the building(s)?   & MC-SS & 880 (0/0) \\ 
    \cdashline{2-2}
    & \textit{(Personally by a responsible person in hallways or classrooms \textbar Using a building } & \\
    & \textit{broadcast \textbar By activating the fire alarm system (e.g., a smoke detector) \textbar Other)} & \\
    \hline
    16 & How are children organized before leaving a classroom after the signal to evacuate is   & MC-SS & 878 (2/0) \\  & given?  & \\
    \cdashline{2-2}
    & \textit{(Children are asked to form pairs and wait until a signal to leave the classroom is given } & \\
    & \textit{(i.e., leaving in pairs as a group) \textbar Children are asked to form a group and wait until a } & \\
    & \textit{signal to leave the classroom is given (i.e., leaving as a group) \textbar  Children are asked to}  & \\
    & \textit{wait until a signal to leave the classroom is given (i.e., without a given formation)} \textbar  & \\
    & \textit{Children can leave a classroom individually \textbar Other)} & \\
    \hline
    17 & How are children organized when moving through the building(s) to a place of safety?    & MC-SS & 878 (2/0) \\  
    \cdashline{2-2}
    & \textit{(Children move in pairs in a compact group (i.e., each class forms a separate group) \textbar } & \\
    & \textit{Children move in a group but not in pairs (i.e., each class forms a separate group)} \textbar  & \\
    & \textit{Children can move individually after being advised to wait in a specific place \textbar Other)} & \\
    \hline
    18 & Are escape routes and exits (e.g.,~external escape staircases) not used daily used during  & MC-SS & 878 (2/0) \\
    & fire drills in your nursery school?  & \\
    \cdashline{2-2}
    & \textit{(Yes \textbar No, children move only using well-known and daily used routes \textbar No, there is only } & \\
    & \textit{one escape route in the building which is also used daily \textbar Other)} & \\
    \hline
    19 & Are fire drills evaluated after they are completed?  & MC-SS & 876 (4/0) \\
    \cdashline{2-2}
    & \textit{(Yes, they are both oral and written evaluations \textbar Yes, oral evaluation \textbar No \textbar Other)}  &  \\ 
    \hline
    20 & What are the most common issues you must resolve during fire drills in your nursery  & MC-MS & 880 (0/0) \\
    & school?  & \\
    \cdashline{2-2}
    &  \textit{(Blocked escape routes (e.g. locked/blocked exits, forgotten keys) \textbar Delays caused by } & \\
    & \textit{waiting for an empty staircase used by multiple classes \textbar Choice of escape route \textbar Delays}  & \\
    & \textit{caused by staff's poor knowledge of evacuation procedures \textbar Slow movement because of}  & \\
    & \textit{children required assistance \textbar Insufficient number of staff present for effectively helping}  & \\
    & \textit{all children requiring assistance \textbar Children are not willing to leave a classroom (e.g.,}  & \\
    & \textit{hiding, running away) \textbar Children do not follow instructions given by staff members \textbar}  & \\
    & \textit{Children are scared or frightened \textbar Other)} & \\
    \hline
    \multicolumn{4}{l}{\# =question number}, MIS=missing response, EXL=invalidated and excluded response \\ 
    \multicolumn{4}{l}{MC-SS=single-select multi-choice question, MC-MS=multi-select multi-choice question} \\
    \hline
    \end{tabular}
\end{table}

\subsection{Data analysis}
\label{subsec:data}
The survey responses were received electronically (automatically recorded in the survey tool), placed in spreadsheets, manually organized and checked for duplicates. Within this process, individual responses were checked for typos and reviewed using the additional information and explanation provided by the respondents in their comments to the survey questions, if applicable. In certain instances, comments provided additional details that allowed us to update the response from the "Other" option to one of the specified categories. The question responses with identified errors (e.g., decimal place mistakes) were invalidated, excluded, and considered missing in the subsequent analysis. The analysis of invalidated responses for each question, as presented in Table~\ref{tab:questions_I} and Table~\ref{tab:questions_II}, reveals that Question \#4 posed the greatest challenge. In total, 27~responses were discarded due to a clear misinterpretation of the question by the participants who reported higher values than 10~teaching staff members present on average in a class. 
Since all questions were on a voluntary basis, several survey responses were received only partially completed, with some questions left unanswered. It was determined that out of total \numprint{1151}~survey responses, \numprint{1068} ($92.8\%$)~survey responses were complete. In 77~survey responses ($6.7\%$) one question response was missing (in 50~cases) or invalidated (in 27~cases) with no discernible pattern apart from disregarded responses for Question \#4. In 6~survey responses ($0.5\%$) 2--4~missing and invalidated question responses were identified. We opted to retain all survey responses with one missing question response, in order not to reduce the sample size and further explored the impact of excluding survey responses with 2--4~missing question responses. Since discarding these survey responses did not cause significant differences in the results, we retain and analyze all incomplete survey responses in this study.
For the numeric questions, the main characteristics of the distribution of the responses were examined, specifically mean, median, and standard deviation as well as the overall shape of the distribution. For categorical questions, the percentages of responses in each category were computed. Depending on their nature, responses in the \quotes{Other} category were either counted as one of the available options or as a separate category. Under-represented categories were merged together for the sake of clarity and computational feasibility (e.g. merging \quotes{No} and \quotes{I don't know} into one category when either was reported only by a small number of respondents). 
We examined the pairwise relation between indicators in Part~I and the fact whether fire drills were performed, as well as the relation between indicators from both parts with the occurrence of specific issues encountered during fire drills and their number. The relationship was evaluated in the following manner:

\begin{itemize}
\item Relations between two numeric indicators were evaluated using the Spearman's correlation coefficient and a significance test was performed.
\item Relations between two categorical indicators were evaluated using Pearson's chi-square independence test in contingency tables.
\item Relations between the numeric and categorical indicators were examined by comparing the means and medians of the numeric variables in individual categories, and a Wilcoxon or Kruskal-Wallis test for equality of medians was performed. 
\end{itemize}

We opted for rank-based non-parametric methods since we could not ensure the normality of the data, with the distributions being discrete and asymmetric. Pairs with statistically significant relations were examined in more detail to determine possible causes. All tests were performed on level of significance $\alpha=5\%$. For more details regarding the methods used see \cite{nonparam_book}. Computations were performed in the R software 4.3.3. \cite{r_core_team_2024}.

\subsection{Limitations}\label{subsec:limit}
Despite the fact that online surveys are useful tools for collecting data, they come with certain limitations such as sample bias, response bias, low motivation of respondents, and technical issues like internet connectivity and knowledge of the required technology \cite{coughlan_survey_2009, topuzovska_latkovikj_online_2020}. In the context of this study, several limitations must be considered in order to properly interpret the results. The survey targeted all officially recorded nursery schools in Czechia (\numprint{5364} nursery schools); however, only \numprint{4930} nursery schools with available email contact were contacted and \numprint{4903} invitation emails were delivered successfully. The survey response rate was of 23.5\% (see Section~\ref{subsec:rate} for details). Non-response bias could arise as the survey could complete respondents with working internet access and those who were sufficiently interested in the topic and willing to give their time \cite{andrade_limitations_2020}. In addition, the willingness to participate in the survey could be impacted by unknown motives. For example, people who did not return the survey may generally be less likely to be used to perform fire drills. 
Surveys are also prone to response bias \cite{atif_estimating_2012}. Respondents can have a tendency to satisfice when answering questions (acquiescence or moderacy bias) and to skew self-reports in a favorable manner (social desirability bias) \cite{furnham_response_1986, bogner_response_2016}. Survey respondents may also have exhibited bias in their responses and comments because of the personal opinions and perceptions \cite{tempelaar_subjective_2020}. 
Another source of response bias could result from the formulation of questions and response alternatives in the survey \cite{bogner_response_2016}. We had to make trade-offs between the use of technical words and the levels of knowledge of potential respondents while designing the survey, a known issue for surveys where target groups not familiar with scientific terminology are involved \cite{andrews_conducting_2003}. Despite efforts to maintain a reasonable balance between being unambiguous and not adding complexity, additional bias could appear due to lexical ambiguity \cite{tourangeau_psychology_2009}.   
Our survey findings could also be constrained by the bias arising from incomplete survey responses, necessitating a deeper investigation into which questions tended to be unanswered, the reasons for this, and the potential impact on the results.
To mitigate survey fatigue \cite{jeong_exhaustive_2023}, the survey instrument was designed with a limited set of questions, which did not cover all aspects of interest for fire safety and evacuation planning. Consequently, a comprehensive examination of all environmental and operational factors affecting fire safety in nursery schools was not possible.

However, the great advantage of online surveys is the possibility of broad dissemination and obtaining a large body of data within a limited amount of time. In our case, this would not have been possible with alternative methods such as structured or semi-structured interviews \cite{topuzovska_latkovikj_online_2020}.

\subsection{Brief context: Preschool education in Czechia}
\label{subsec:edu}
Preschool education in Czechia is provided by nursery schools and is generally intended for children between 3~and 6~years of age (between 2~and 7~years, in exceptional cases). Preschool education is compulsory only for children in the pre-primary school year (that is, for children at least 5~years of age). Nursery schools are organized into classes of up to 28~children. In terms of the age, classes can be homogeneous (children of similar age in one class) or heterogeneous (children with different ages in one class). Children with special educational needs can be integrated into standard classes.\footnote{If support measures provided in a school are not sufficient to meet educational needs, children with special educational needs or disabilities (such as mental, visual, hearing, physical disabilities, serious developmental learning impairments, speech impediments, multiple impairment, autism), these children can be educated separately in special classes or in independent nursery schools catering to special needs children.} Children in nursery schools can attend full-time or part-time based on the legal guardians' decision and educational program available. The main educational requirements and rules are determined by an official framework; however, nursery schools can have their own educational programs including some alternatives (e.g. Montessori, Waldorf, or Dalton methods). Children are taught by teachers specialized in preschool education, mainly graduates of university-level education schools \cite{noauthor_act_2004, doe_early_2017}.

\section{Results} 
\label{sec:results}
This section begins by evaluating the survey response rate (Section~\ref{subsec:rate}) and then presenting the results with descriptive analysis (Section~\ref{subsec:descriptive}) and statistical analysis (Section~\ref{subsec:statistic}). Indicators used for interpretation of the results are named in Table~\ref{tab:partI} and Table~\ref{tab:partII}. For clarity, these indicators are linked to survey questions listed in Table~\ref{tab:questions_I} and Table~\ref{tab:questions_II} (\#2--20). It should be noted that the term \quotes{staff members} refers solely to \quotes{teaching staff} (e.g., teachers, teacher assistants, educators). 

\subsection{Response rate} \label{subsec:rate}
\begin{figure}[t]
\centering
\includegraphics[width=0.75\textwidth]{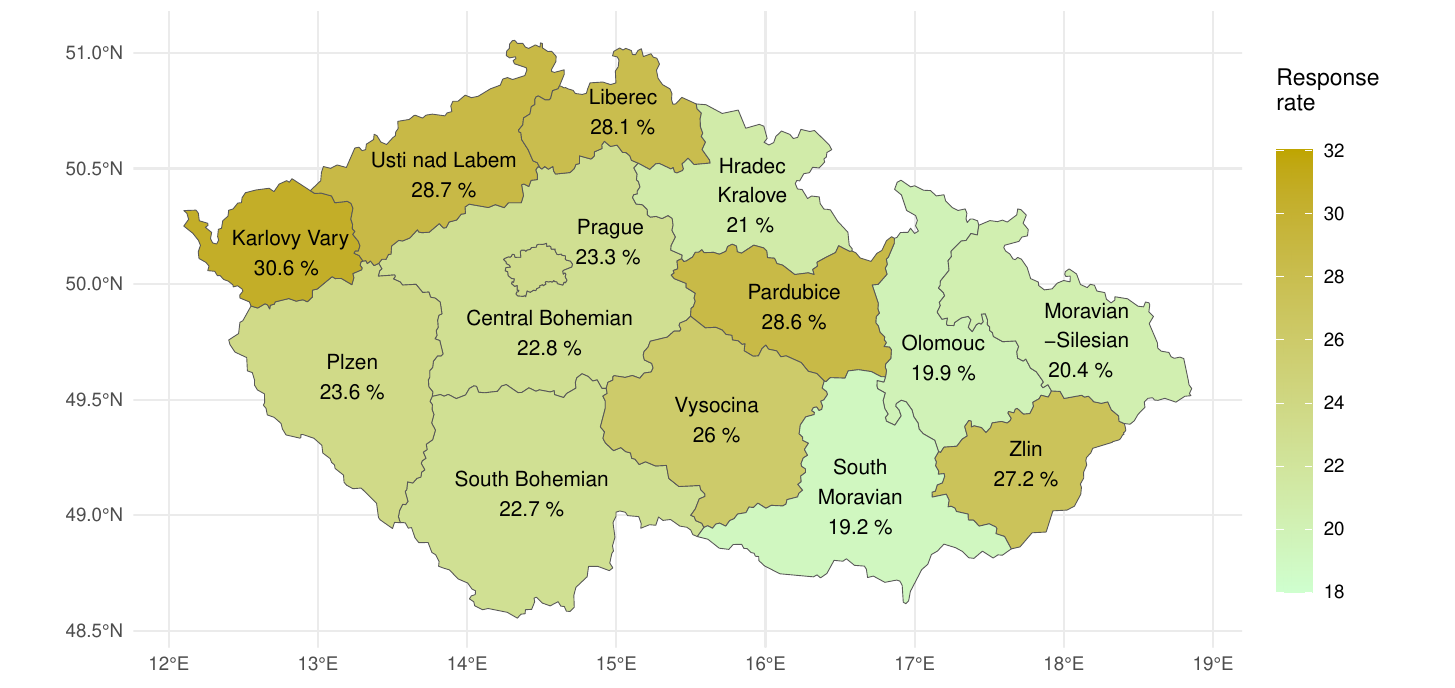}
\caption{Response rate for each Czech region}
\label{fig:regions}
\end{figure}
 
\numprint{1151} survey responses represented a response rate of 23.5\%. The response rate was calculated as the relationship between the number of survey responses returned and the number of invitations delivered. The survey responses exhibited a fairly balanced geographical coverage of Czechia, with the number of respondents for each region ranging from 19.2\% to 30.6\% (Fig.~\ref{fig:regions}). The response rate in regions was calculated as the ratio between the number of survey responses in a region and the number of invitations delivered in a region. \numprint{1151} survey respondents represented 21.5\% of all officially registered nursery schools in Czechia.
Several surveys were only partially completed, with some questions left unanswered (see Table~\ref{tab:questions_I} and Table~\ref{tab:questions_II} for the number of responses received for each survey question). In total \numprint{1472}~comments were received (378~comments for Part~I and \numprint{1094}~comments for Part~II) and were evaluated in detail. When applicable these where integrated into the interpretation of results. \par

\subsection{Descriptive analysis}
\label{subsec:descriptive}
\subsubsection{Part~I: General information} \label{subsubsec:PartI}
\begin{figure}[h!]
\centering
\includegraphics[width=\textwidth]{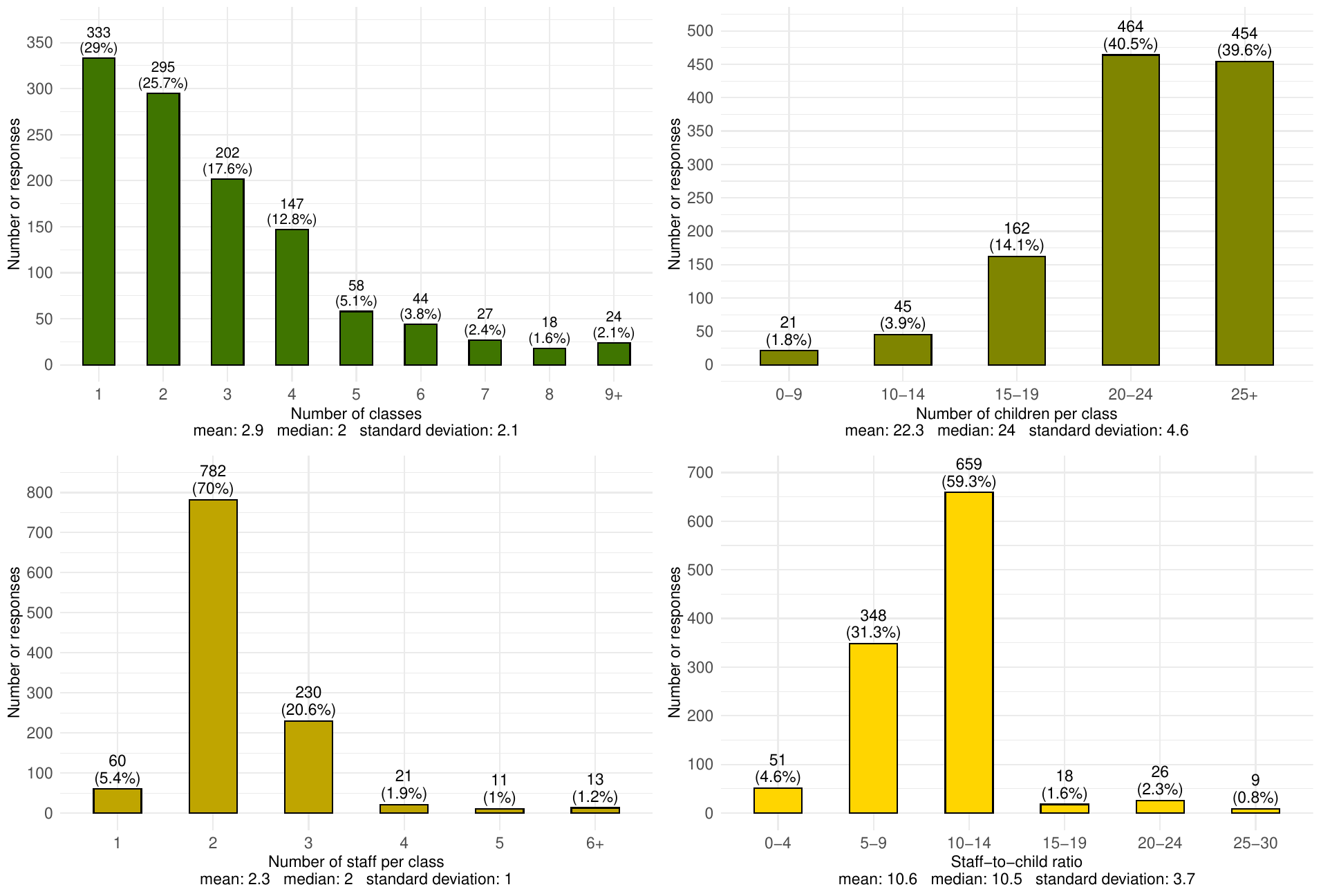}
\caption{Capacity and occupancy of nursery schools in this study. \textbf{Top left:} Number of classes in the nursery school (\#2); \textbf{Top right:} Number of children in a class (\#3); \textbf{Bottom left:} Number of staff in a class (\#4); \textbf{Bottom right:} Staff-to-child ratio}
\label{fig:classesstaff}
\end{figure}

This section briefly introduces the results of Part~I of the survey (n=\numprint{1151}~responses) to provide insight into the specifics of the nursery school environments probed in this study. The first section focused on occupancies (e.g.,~number of persons assumed present and staff-to-child ratios) using numeric questions; results are presented in Figure~\ref{fig:classesstaff}.
One-class nursery schools were the most common type of school, followed by nursery schools with 2~to 4~classes (Figure~\ref{fig:classesstaff} Top left). Nursery schools with more than 4~classes were less common; however, it can be assumed that nursery schools with 5~to 9~classes might be still often located in one building.\footnote{Note that nursery schools as legal entities may have multiple buildings or workplaces, which may be why some respondents reported class values.}  The most frequent number of children in a class was between 20~and 28~children (80.1\% of respondents) with a median value of 24~children, \footnote{Note that both meanings, the average number of enrolled children in a class and the average number of children usually present in a class must be considered when evaluating responses.} see Figure~\ref{fig:classesstaff} Top right. Most nursery schools reported two staff members per class, followed by 3~staff members in a class (Figure~\ref{fig:classesstaff} Bottom left). However, comments made by respondents revealed that more than one staff member in a class might be the case for only part of the day due to staggered shifts (typically several hours around noon). Different numbers of staff members should be assumed for children with special educational needs classes and classes in special nursery schools as well as classes intended for toddlers and younger preschool children. In those cases, the average number of children per class was between 6~and 10~children, and the average number of staff members in a class might be considered higher than in normal classes, e.g.,~3--4~staff with fewer children present. We calculated the staff-to-child ratio separately for each respondent as the ratio between the number of children and the number of staff members in a class (Figure~\ref{fig:classesstaff} Bottom right). The most frequently calculated staff-to-child ratios were 10~to 14~children per staff member and between 5~and 9~children per staff member. More than 15~children per staff member occurred only in 4.7\% of cases.

\begingroup
  \renewcommand{\arraystretch}{0.7}  
\begin{table}[hbt!]
\centering
\footnotesize
\caption{Summary of results, Part~I ($n=1151$ responses in total). Indicators are linked to particular survey questions (\#5--10); the number of question responses is provided separately for each indicator}
\label{tab:partI}
\begin{tabular}{R{7.5cm}|C{2cm}|C{2.5cm}}
\textbf{Indicator} & \textbf{Responses} & \textbf{Frequency [\%]}  \\
\hline
\multicolumn{3}{l}{\textbf{Organization of classes according to children's ages} (\#5, 1144~responses in total)} \\
\hline
\hline
Heterogeneous classes & 650 & 56.8 \\
Homogeneous classes & 428 & 37.4 \\
Combination & 66 & 5.8 \\
\hline
\multicolumn{3}{l}{\textbf{Maximum level location of spaces intended for children} (\#6, 1151~responses in total)} \\
\hline
\hline
Ground floor & 387 & 33.6 \\
First floor & 632 & 54.9 \\
Higher than first floor & 132 & 11.5 \\
\hline
\multicolumn{3}{l}{\textbf{External escape staircase(s) available} (\#7, 1143~responses in total)} \\
\hline
\hline
Yes  & 403 & 35.3 \\
No & 740 & 64.7 \\
\hline
\multicolumn{3}{l}{\textbf{Fire safety documentation available} (\#8, 1144~responses in total)} \\
\hline
\hline
Yes  & 1067 & 93.3 \\
No  & 36 & 3.1 \\
Unknown & 41 & 3.6 \\
\hline
\multicolumn{3}{l}{\textbf{Educational program includes personal safety} (\#9, 1144~responses in total)} \\
\hline
\hline
Yes  & 1135 & 99.2 \\
No/Unknown   & 9 & 0.8 \\
\hline
\multicolumn{3}{l}{\textbf{Fire drills performed} (\#10, 1151~responses in total)} \\
\hline
\hline
Yes  & 880 & 76.5 \\
No   & 271 & 23.5 \\
\hline
\end{tabular}
\end{table}
\endgroup

The results described with categorical indicators are summarised in Table~\ref{tab:partI}. Considering the organization of classes in terms of children’s ages, in more than half of the cases, nursery schools responded that children are organized in heterogeneous classes (i.e.~standard age range 3~to 6~years). For homogeneous classes, there were usually 2~or 3~different age groups per school (87.7\%). However, the respondents also reported both options in the \quotes{Other} category, indicating a combination of heterogeneous classes (e.g.,~children 3--5~years) and separate primary school preparatory classes (e.g.,~children aged 5--7~years), or separate classes for very young children (e.g.,~2--3~years).
Regarding the buildings in which the nursery schools were located (specifically, which floors the spaces intended for children such as classrooms, lunchrooms, sleeping rooms, gyms were located on), and whether external escape staircases were available in the buildings, most respondents reported that spaces are on the ground floor and/or on the first floor (88.5\% of respondents). Comments revealed that elevated ground floors are also common and staircase-free escape routes cannot be automatically assumed for all ground floor locations. External escape staircases were present in 35\% of nursery schools. Schools located only on the ground floor were equipped with external escape staircases in 26.4\% of cases and schools situated on the ground floor or first floor in 39.2\% of cases. Among schools on higher floors, external escape staircases were present in 40.2\% of cases. 880~nursery schools (76.5\%) responded that they have fire drills and were also invited to participate in Part~II of the survey. \par

\subsubsection{Part~II: Fire drills} \label{subsubsec:PartII}
Part~II was only accessible by nursery schools who reported having fire drills. This part focused on details related to the drills such as frequency, level of information provided to participants, warning signals, escape routes used, how children are organized in drills, issues experienced, and if drills are evaluated after a drill. Results from Part~II (n=\numprint{880}~responses) are summarized in Table~\ref{tab:partII} and interpreted below. \par
According to respondents, the majority of schools (93\%) performed fire drills annually (82.9\%) or biannually (10.1\%).\footnote{At least once in the last 5~years.} Fire drills were announced to staff in some detail in 84\% of schools.\footnote{A frequent argument for announced fire drills in comments was preventing children's stress and anxiety.} The most common way that staff learned about fire drill behavior (78.5\% of responses) was regularly-occurring fire protection education provided by a fire prevention officer (i.e.,~a fire prevention professional), sometimes combined with internal education provided by school's management (combined in 24.4\% of responses).
Oral warning signals (for example, saying \quotes{Fire}, \quotes{Fire alarm}, \quotes{Fire drill}, or \quotes{Evacuation}) were given in 35.5\% of schools. Manual sound signals (banging on a metal object, whistle, bell, siren, or gong) were used in 23.4\% of cases. A combined oral command and manual signal was used in 37.8\% of cases. In most nursery schools (88.1\%), the warning signal for starting evacuation was given personally by a responsible person in corridors or classrooms.

\begingroup
  \renewcommand{\arraystretch}{0.7}  
\begin{table}[hbt!]
\centering
\footnotesize
\caption{Summary of results, Part~II ($n=880$ responses). Indicators are linked to particular survey questions (\#11--19); the number of question responses is provided separately for each indicator}
\label{tab:partII}
\begin{tabular}{R{7.5cm}|C{2cm}|C{2.5cm}}
\textbf{Indicator} & \textbf{Responses} & \textbf{Frequency [\%]}  \\
\hline
\multicolumn{3}{l}{\textbf{Frequency of fire drills} (\#11, 872~responses in total)} \\
\hline
\hline
Less than 1$\times$ per 2~years   & 45 & 5.2 \\
1$\times$ per 2~years & 13 & 1.5 \\
Annually & 723 & 82.9 \\
Biannually & 88 & 10.1 \\
More than 2$\times$ per year & 3 & 0.3 \\
\hline
\multicolumn{3}{l}{\textbf{Announcement of fire drills to staff} (\#12, 880~responses in total)} \\
\hline
\hline
Announced with date and time in advance & 373 & 42.4 \\
Announced with date in advance & 182 & 20.7 \\
Announced in advance but without date and time  & 184 & 20.9 \\
Unannounced & 141 & 16 \\
\hline
\multicolumn{3}{l}{\textbf{Information and instruction source for staff} (\#13, 880~responses in total)} \\
\hline
\hline
Regular fire protection education    & 473 & 53.8 \\
Internal education                   & 168 & 16.1 \\
Both previous options		         & 215 & 24.4 \\
No special education                 & 21 &  2.4 \\
Other                                & 3  &  0.3 \\
\hline
\multicolumn{3}{l}{\textbf{Type of warning signal} (\#14, 876~responses in total)} \\
\hline
\hline
Oral command & 311 & 35.5 \\
Manual sound signal & 205 & 23.4 \\
Oral command and manual sound signal in combination & 331 & 37.8 \\
Fire alarm system & 21 & 2.4 \\
Coded message or other & 8 & 0.9 \\
\hline
\multicolumn{3}{l}{\textbf{Warning signal distribution} (\#15, 880~responses in total)} \\
\hline
\hline
Personally by staff members &  775 & 88.1 \\
Broadcast & 42 & 4.8 \\
Fire alarm system & 26 & 3 \\
Other & 37 & 4.2 \\
\hline
\multicolumn{3}{l}{\textbf{Exit strategy during the pre-movement phase} (\#16, 878~responses in total)} \\
\hline
\hline
Waiting for signal as a group in pairs & 464 & 52.8  \\
Waiting for signal as a group  & 263 & 30 \\
Waiting for signal without formation & 131 & 14.9 \\
Children free to leave on their own & 6 & 0.7 \\
Other & 14 & 1.6 \\
\hline
\multicolumn{3}{l}{\textbf{Organization during the movement phase} (\#17, 878~responses in total)} \\
\hline
\hline
As a group in pairs & 438 & 49.9  \\
As a group  & 386 & 44 \\
Individually & 43 &  4.9 \\
Other & 11 &  1.3 \\
\hline
\multicolumn{3}{l}{\textbf{Usage of escape routes} (\#18, 878~responses in total)} \\
\hline
\hline
Escape routes not used daily employed & 361 & 41.4  \\
Only known escape routes used  & 337 & 38.4 \\
Only one route available & 164  & 18.7  \\
Other & 16 & 1.8 \\
\hline
\multicolumn{3}{l}{\textbf{Evaluation after fire drill completed} (\#19, 876~responses in total)} \\
\hline
\hline
Oral and written form & 540 & 61.6  \\
Oral form & 312 & 35.6 \\
No evaluation & 24  & 2.7  \\
\hline
\end{tabular}
\end{table}
\endgroup

\begin{figure}[h!]
\centering
\includegraphics[width=\textwidth]{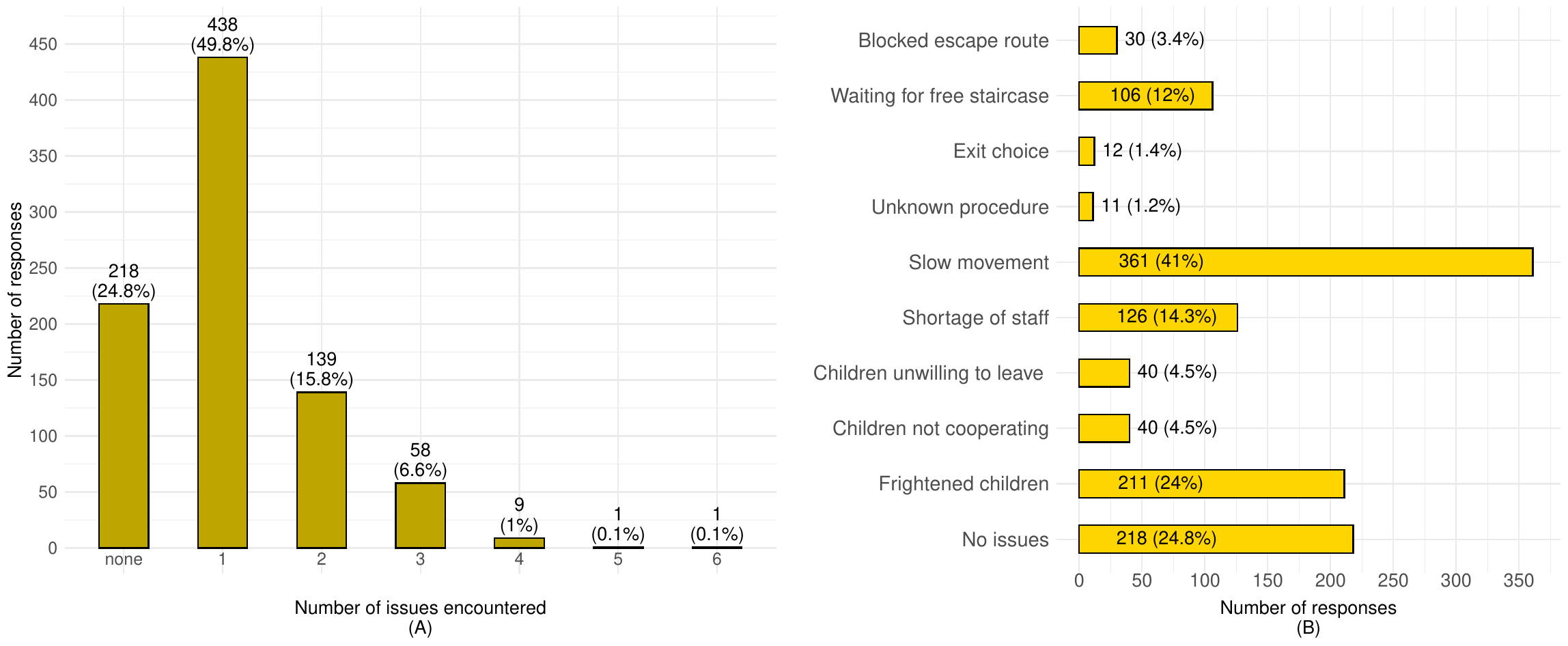}
\caption{Issues encountered during fire drills in nursery schools: (A) Number of issues; (B) Frequency of various issues reported by survey respondents}
\label{fig:issues}
\end{figure}

Regarding strategies on how to organize children during fire drills before exiting a classroom, 97.7\% of respondents noted children were asked to wait for further instructions before exiting. When traveling through their schools to an exit, children were organized in groups by class in 93.9\% of schools. In nearly half (49.9\%) of those cases, children were asked to form pairs (some groups of three) and hold hands. Based on comments, organized exiting in single file was also a common strategy. 
Routes usually taken daily (well known to fire drill participants) were used as escape routes according to 57.1\% of respondents. 
Fire drills were usually followed by debriefing according to 97.2\% of respondent (in 61.6\% of cases, in written form). Respondents revealed that total evacuation time is commonly measured as part of the evaluation process. However, the survey did not probe to what extent fire drill evaluations are performed.
Regarding issues experienced during fire drills, 73.4\% of respondents said, at least one issue occurred during previous fire drills (see Figure~\ref{fig:issues} (A)). The most frequently encountered complication (according to 41\% of respondents) was slow movement because children required assistance. Respondents also noted that frightened children were an issue (24\%)\footnote{Note that comments regarding anxiety and fear in children during fire drills were entirely based on views of the respondents.} as well as an insufficient number of staff who could effectively help all the children who required assistance (14.3\%) and delays caused by waiting for a staircase to be free (14\%), see Figure~\ref{fig:issues} (B). Based on comments from respondents, younger children (generally those under 4~years of age) needed more assistance (particularly on stairs) and showed more frequent signs of anxiety during fire drills. \par

\subsection{Statistical analysis}
\label{subsec:statistic}
In addition to descriptive analysis of results, we conducted statistical analysis to further explore our findings. Statistical analysis focused on three main areas: Which nursery schools conducted fire drills (Section~\ref{subsubsec:where}), which nursery schools were more likely to encounter issues during fire drills (Section~\ref{subsubsec:issues}), and what types of issues were most common (Section~\ref{subsubsec:specific}). 

\subsubsection{Conducting vs. not conducting fire drills} \label{subsubsec:where} 
\begin{figure}[h!]
\centering
\includegraphics[width=0.8\textwidth]{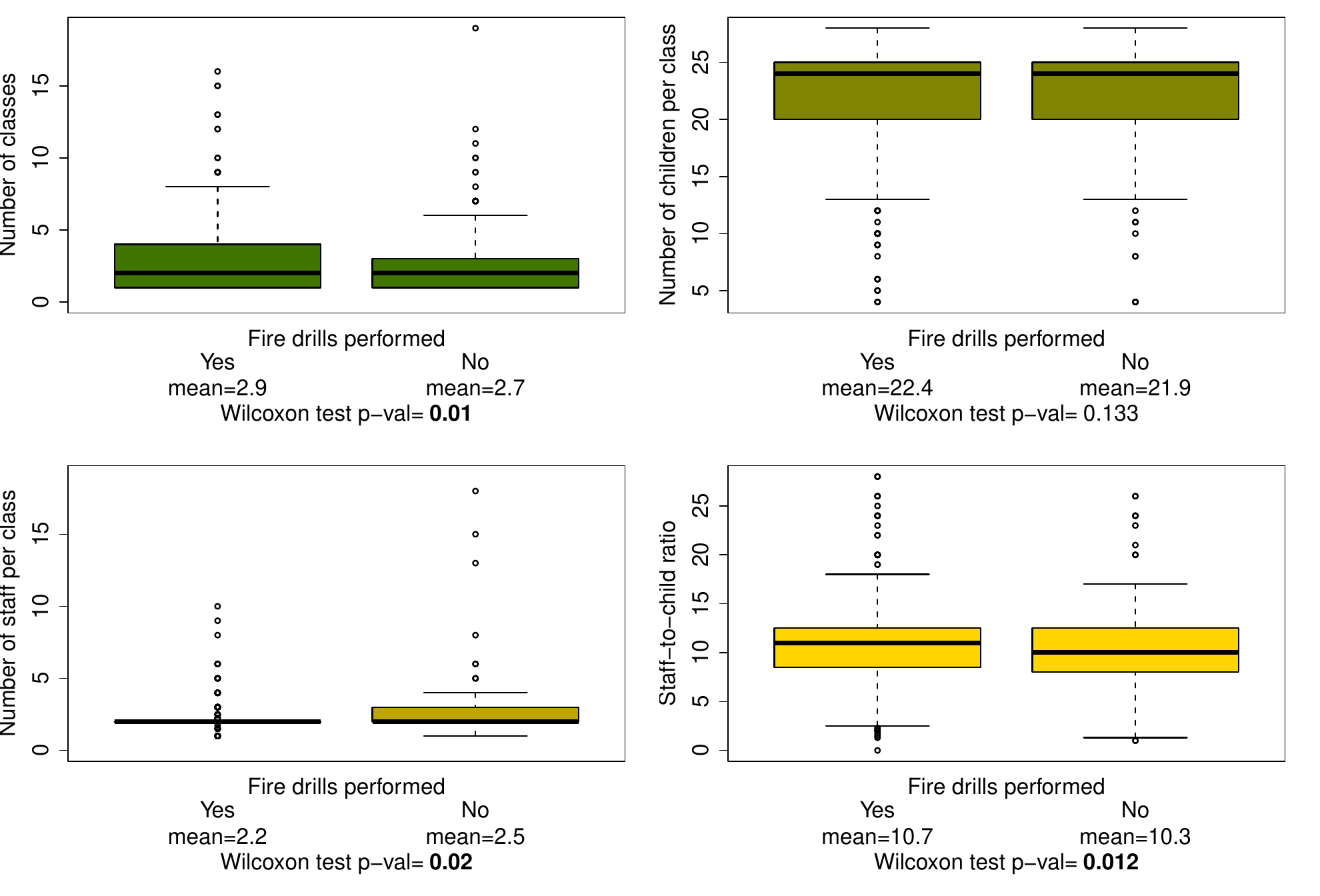}
\caption{Comparing numeric indicators for schools which do and do not perform drills (significant $p$-values in bold). \textbf{Top left:} Number of classes in a nursery school; \textbf{Top right:} Number of children in a class; \textbf{Bottom left:} Number of staff members in a class; \textbf{Bottom right:} Staff-to-child ratio}.
\label{fig:evacperc_num22}
\end{figure}

To explore factors that are of relevance to fire safety practices and that might affect whether fire drills are executed, our first objective was to analyze responses in Part~I to determine what differentiates nursery schools that performed fire drills (880~responses) from those that did not (remaining 271~responses). First, we used numeric indicators and compared the mean of the number of classes, the number of children in a class, the number of staff members in a class, and the staff-to-child ratio for schools that perform fire drills as opposed to those who do not. We tested their equality using a two-sample Wilcoxon test. While the indicators are closely related, they each provide different insight. As seen in Figure \ref{fig:evacperc_num22}, nursery schools that performed drills tended to have a significantly higher number of classes (2.9 versus 2.7, $p$-val=0.01), significantly fewer staff in a class (2.2 versus 2.5, $p$-val=0.02), and a significantly higher staff-to-child ratio (10.7 versus 10.3, $p$-val=0.012). In contrast, the difference between the mean number of children per class between the schools that performed fire drills and those that did not is not statistically significant (22.4 versus 21.9, $p$-val=0.133). 

To observe the percentage of nursery schools that conducted drills according to categorical descriptors, the Chi-Square independence test was performed (see Table~\ref{tab:evacperc_cat}). No association was found for how children were placed in age-based classes (homogeneous versus heterogeneous versus combined classes) or for the location of spaces intended for children in a school. Surprisingly, in our analysis, whether a school had an external escape staircase(s) or not was also not relevant. 

\begin{table}[hbt!]
\centering
\footnotesize
\caption{Percentage of nursery schools performing fire drills depending on categorical indicators}
\label{tab:evacperc_cat}
\begin{tabular}{R{4cm}|C{5cm}|C{2.5cm}}
\textbf{Indicator} & \textbf{Schools performing fire drills [\%]} & \textbf{$p$-val [\%]}  \\
\hline
\multicolumn{3}{l}{\textbf{Organization of classes according to children's ages} (\#5)} \\
\hline
\hline
Heterogeneous classes & 74.8 & \multirow[c]{3}{*}{0.245} \\
Homogeneous classes  & 78.3 &  \\
Combination   & 80.3 &  \\
\hline
\multicolumn{3}{l}{\textbf{Maximum level location of spaces intended for children} (\#6)} \\
\hline
\hline
Ground floor  & 76.5 & \multirow[c]{3}{*}{0.652}  \\
First floor   & 75.8 & \\
Higher than first floor      & 89.5 & \\
\hline
\multicolumn{3}{l}{\textbf{External escape staircase(s) available} (\#7)} \\
\hline
\hline
Yes  & 79.7 & \multirow[c]{2}{*}{0.072} \\
No   & 74.7 & \\
\hline
\end{tabular}
\end{table}

It should be noted that the methods for statistical analysis (Spearman's and Chi-square tests) were chosen to allow the study of unbalanced samples (since in our analysis the number of schools that performed fire drills was larger than of those that did not).

\subsubsection{Number of issues encountered } \label{subsubsec:issues} 
Further analysis focused on relationships between the number of issues respondents reported (see also predefined list in Table~\ref{tab:questions_II} and Figure~\ref{fig:issues}) and the indicators from both Part~I and Part~II of the survey.
First, we explored the indicators from Part~I that characterized the schools and operational conditions in them. The Spearman's correlation coefficient was calculated between the number of issues encountered and the number of classes, the number of children in a class, the number of staff in a class, and the staff-to-child ratio. After this, Spearman's test for non-correlation was performed. Results are summarized in Table~\ref{tab:statissueno_num}. The correlation coefficients were rather small, yet a significantly positive correlation was found at 5\% level of significance between the number of issues encountered and the number of classes ($p$-val=0.028); however, this may be only a computational artifact arising from the fact that both the number of issues and number of classes have only whole numbers as values and the $p$-values produced by the Spearman's test are not exact.

\begin{table}[h!]
\centering
\footnotesize
\caption{Correlation of the number of issues encountered during fire  drills with numeric indicators (significant $p$-value in bold)}
\label{tab:statissueno_num}
\begin{tabular}{R{5cm}|C{5cm}|C{2.5cm}}
\textbf{Indicator} & \textbf{Correlation with no. of issues} & \textbf{$p$-val} \\ \hline
Number of classes (\#2)			  & 0.074 & \textbf{0.028} \\
Number of children per class (\#3)  & -0.026  & 0.473 \\
Number of staff per class (\#4)	  & -0.020  & 0.518 \\
Staff-to-child ratio		              & 0.004  & 0.907 \\ \hline
\end{tabular}
\end{table}

Significance of the number of issues encountered for individual levels of categorical indicators from both Part~I and Part~II of the survey was tested using the two-sample Wilcoxon test for dichotomic categories and the Kruskal-Wallis test for multicategorical indicators. The mean number of issues per category and the $p$-values of the tests can be found in Table \ref{tab:statissueno_cat}. Nursery schools with rooms intended for children only on the ground floor reported fewer problems (0.94) on average than schools with rooms on the first floor (1.10) or higher floors (1.26), with a $p$-val=0.010. Furthermore, nursery schools that conduct fire drills less than once a year encountered significantly more problems (1.45, $p$-val=0.023) than schools that organize annual (1.03) or more frequent drills (1.01). Other indicators were not significantly related to the number of issues encountered. 

\begin{table}[h!]
\centering
\footnotesize
\caption{Comparing the mean number of issues encountered during fire drills in nursery schools depending on categorical indicators from Part~I (significant $p$-value in bold)}
\label{tab:statissueno_cat}
\begin{tabular}{R{5cm}|C{4cm}|C{2.5cm}}
\textbf{Indicator} & \textbf{Mean} no. of issues & \textbf{$p$-val} \\ \hline

\multicolumn{3}{l}{\textbf{Organization of classes according to children’s ages (\#5)} } \\
\hline
\hline
Heterogeneous classes & 1.03 & \multirow[c]{3}{*}{0.086} \\
Homogeneous classes   & 1.06 &  \\
Combination   & 1.34 &  \\
\hline
\multicolumn{3}{l}{\textbf{Maximum level location of spaces intended for children (\#6)}} \\
\hline
\hline
Ground floor  & 0.94 & \multirow[c]{3}{*}{\textbf{0.010}}  \\
First floor   & 1.10 & \\
Higher than first floor        & 1.26 & \\
\hline
\multicolumn{3}{l}{\textbf{External escape staircase(s) available (\#7)}} \\
\hline
\hline
Yes & 1.03 & \multirow[c]{2}{*}{0.541} \\
No   & 1.09 & \\
\hline
\end{tabular}
\end{table}

\renewcommand{\arraystretch}{0.7}   
\begin{table}[h!]
\centering
\footnotesize
\caption{Comparing the mean number of issues encountered during fire drills in nursery schools depending on categorical indicators from Part~II (significant $p$-value in bold)}
\label{tab:statissueno_num3}
\begin{tabular}{R{5cm}|C{4cm}|C{2.5cm}}
\textbf{Indicator} & \textbf{Mean} no. of issues & \textbf{$p$-val} \\ 
\hline
\multicolumn{3}{l}{\textbf{Frequency of fire drills (\#11)}} \\
\hline
\hline
Less often & 1.45 & \multirow[c]{3}{*}{\textbf{0.023}} \\
Annually   & 1.03 & \\
More often & 1.01 & \\
\hline
\multicolumn{3}{l}{\textbf{Announcement of fire drills to staff (\#12)}} \\
\hline
\hline
A: With date and time   & 1.11 & \multirow[c]{4}{*}{0.120} \\
A: With date & 1.07 & \\
A: Without date and time  & 1.11 & \\
Unannounced & 0.88 & \\
\hline
\multicolumn{3}{l}{\textbf{Information and instruction source for staff (\#13)}} \\
\hline
\hline
Regular fire protection education & 1.10 & \multirow[c]{4}{*}{0.251} \\
Internal education  & 0.93 & \\
Both previous options		  & 1.11 & \\
No special education or other & 1.10 & \\
\hline
\multicolumn{3}{l}{\textbf{Type of warning signal (\#14) }} \\
\hline
\hline
Oral command & 1.02 & \multirow[c]{5}{*}{0.195}\\
Manual sound signal & 1.05 & \\
Oral command and sound signal & 1.08 & \\
Fire alarm system & 1.19 & \\
Coded message or other & 1.75 & \\
\hline
\hline
\multicolumn{3}{l}{\textbf{Warning signal distribution (\#15)}} \\
\hline
\hline
Personally by staff &  1.08 & \multirow[c]{4}{*}{0.098}\\
Broadcast & 0.9 & \\
Fire alarm system & 1.12 & \\
Other & 0.78 & \\
\hline
\multicolumn{3}{l}{\textbf{Exiting strategy during the pre-movement phase (\#16)}} \\
\hline
\hline
Waiting as a group in pairs & 1.05 & \multirow[c]{5}{*}{0.933}\\
Waiting as a group  & 1.08 & \\
Waiting without group formation & 1.10 & \\
Children leave on their own & 1 & \\
Other & 1 & \\
\hline
\multicolumn{3}{l}{\textbf{Organization during the movement phase (\#17)}} \\
\hline
\hline
As a group in pairs & 1.04 & \multirow[c]{4}{*}{0.326}\\
As a group  & 1.09 & \\
Individually & 0.98 & \\
Other & 1.64 & \\
\hline
\multicolumn{3}{l}{\textbf{Usage of escape routes (\#18)}} \\
\hline
\hline
Escape routes not daily used & 1.03 & \multirow[c]{4}{*}{0.101}\\
Only known escape routes & 1.10 & \\
Only one route available & 1.10 & \\
Other & 0.62 & \\
\hline
\multicolumn{3}{l}{\textbf{Evaluation after fire drill completed (\#19) }} \\
\hline
\hline
Oral and written form & 1.03 & \multirow[c]{3}{*}{0.192}\\
Oral form & 1.13 & \\
No evaluation & 0.95  &  \\
\hline
\end{tabular}
\end{table}

\subsubsection{Specific issues encountered} \label{subsubsec:specific} 
The most frequently encountered issues reported according to respondents were \textit{Slow movement} (41\% respondents), \textit{Frightened children} (24.8\%), \textit{Shortage of staff} (14.3\%), and \textit{Waiting for free staircase} (12\%); see Table~\ref{tab:questions_II} for the full wording of predefined issues. We focused on examining the relation between these issues and indicators reported in both Part~I and Part~II of the survey.
Table~\ref{tab:statissuetype_num} shows that schools in which evacuation processes were considerably slowed because children required assistance reported a significantly smaller number of children per class ($p$-val=0.022), a larger number of staff per class ($p$-val=0.006) and a smaller staff-to-child ratio ($p$-val$<$0.001). An insufficient number of staff for helping all children requiring assistance (\textit{Shortage of staff}) was reported as an issue by schools with a significantly higher number of classes ($p$-val=0.034), a higher number of children per class ($p$-val=0.027), a lower number of staff per class ($p$-val=0.006), and a higher staff-to-child ratio ($p$-val$<$0.001), which was actually expected. Schools with multiple classes that reported delays caused by waiting for a staircase to be free (\textit{Waiting for free staircase}) had a significantly higher number of classes ($p$-val$<$0.001) and a higher staff-to-child ratio ($p$-val=0.013). The issue of frightened children was not statistically significantly related with numerical indicators.

\begin{table}[h!]
\centering
\footnotesize
\caption{Comparing the means of numeric indicators for nursery schools that did and did not encounter specific issues (significant $p$-values of the Wilcoxon test in bold)}
\label{tab:statissuetype_num}

\begin{tabular}{ll|cc|cc|cc|cc}
&& \multicolumn{2}{c|}{Slow} & \multicolumn{2}{c|}{Frightened} & \multicolumn{2}{c|}{Shortage} & \multicolumn{2}{c}{Waiting for} \\ 
&& \multicolumn{2}{c|}{movement} & \multicolumn{2}{c|}{children} & \multicolumn{2}{c|}{of staff} & \multicolumn{2}{c}{free staircase} \\ \hline
\textbf{Indicator} && \textbf{Yes} & \textbf{No} & \textbf{Yes} & \textbf{No} & \textbf{Yes} & \textbf{No} & \textbf{Yes} & \textbf{No} \\ \hline \hline
\multirow{2}{*}{Number of classes (\#2)} & mean 			& 2.76 & 3.00 & 2.98 & 2.88 & 3.21 & 2.85 & 3.64 & 2.80 \\ \cline{2-10}
&$p$-val	& \multicolumn{2}{c|}{0.157} & \multicolumn{2}{c|}{0.341} & \multicolumn{2}{c|}{\textbf{0.034}}& \multicolumn{2}{c}{\textbf{$<$0.001}} \\ \hline\hline
\multirow{2}{*}{Number of children per class (\#3)}	& mean& 21.9 & 22.7 & 22.4 & 22.4 & 23.14 & 22.26 & 23.08 & 22.29 \\ \cline{2-10}
&$p$-val	& \multicolumn{2}{c|}{\textbf{0.022}} & \multicolumn{2}{c|}{0.753} & \multicolumn{2}{c|}{\textbf{0.027}}& \multicolumn{2}{c}{0.315} \\ \hline\hline
\multirow{2}{*}{Number of staff per class (\#4)} & mean	& 2.32  & 2.19 & 2.22 & 2.25 & 2.05 & 2.28 & 2.16 & 2.26 \\ \cline{2-10}
&$p$-val	& \multicolumn{2}{c|}{\textbf{0.006}} & \multicolumn{2}{c|}{0.228} & \multicolumn{2}{c|}{\textbf{0.003}}& \multicolumn{2}{c}{0.097} \\ \hline\hline
\multirow{2}{*}{Staff-to-child ratio}& mean		& 10.3  & 11.1 & 10.8 & 10.7 & 11.8 & 10.6 & 11.7 & 10.6 \\ \cline{2-10} 
&$p$-val	& \multicolumn{2}{c|}{\textbf{$<$0.001}} & \multicolumn{2}{c|}{0.674} & \multicolumn{2}{c|}{\textbf{$<$0.001}}& \multicolumn{2}{c}{\textbf{0.013}} \\ \hline\hline
\end{tabular}
\end{table}

\begin{figure}[h!]
\centering
\includegraphics[width=\textwidth]{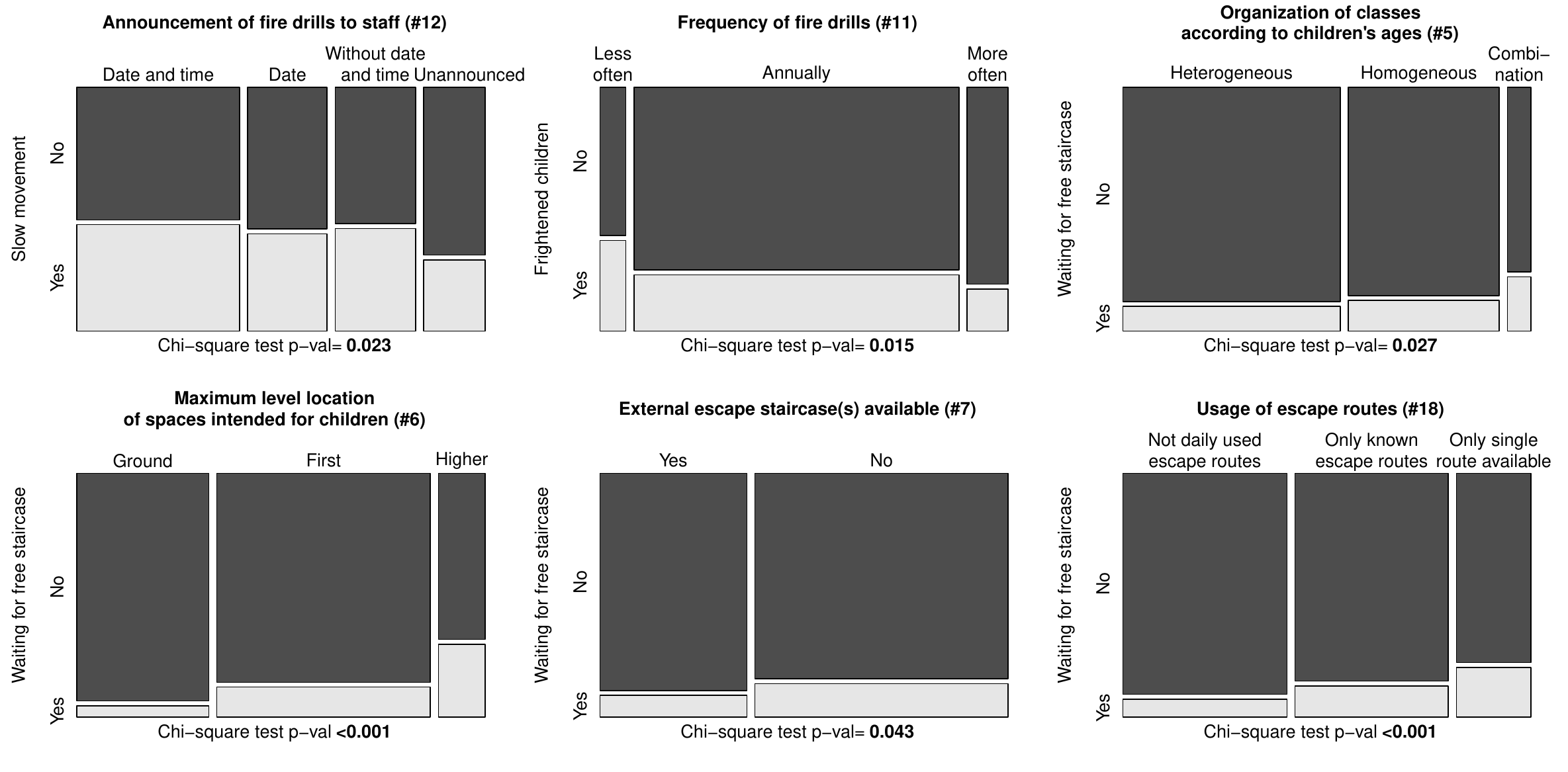}
\caption{Proportion of nursery schools within specific categories that encounter issues  (significant relations only)}
\label{fig:stats_issues_mosaic}
\end{figure}

Table~\ref{tab:probchisq} shows the results for the association between categorical indicators and specific issues\footnote{To satisfy the assumptions of the chi-square test, at least five expected occurrences are needed in each category. For this purpose, the categories with small representation were merged together and the categories labeled as \quotes{Other} were disregarded when under-represented.} statistically significant relations are shown in Figure~\ref{fig:stats_issues_mosaic}. The percentage of nursery schools that experienced slow evacuation movement because children needed assistance varied significantly depending on whether fire drills were announced to staff beforehand ($p$-val=0.023). This issue was encountered less often when fire drills were not announced (29.8\% cases) than when the date/date and time were known beforehand (40.7\% and 44.5\% respectively) or when a drill announcement did not have a specific date (42.9\%). Our results showed that children were more likely to be distressed the less frequently fire drills were performed: being frightene was an issue in 37.9\% of cases, if drills were performed once every two years or less; 23.5\%, when performed annually; and 17.6\% when organized more frequently ($p$-val=0.015). Delays caused by the need to wait for a free staircase were statistically significantly more often observed: 1) for schools that accommodated a mixture of homogeneous and heterogeneous age classes (22.6\% of respondents), in contrast to solely heterogeneous or homogeneous classes (10.3\% and 12.8\% respectively) with $p$-val=0.027; 2) in schools located in buildings with two or more floors (29.8\%) rather than schools with rooms intended for children on the ground floor (4.7\%) or the first floor (12.5\%), with $p$-val$<0.001$; 3) in schools with no external escape staircase(s) available (13.9\%; 9\% when an escape staircase(s) was available), with $p$-val=0.043; and 4) when there was only one escape route and also circulation route available in the school (20.7\%) compared to cases where dedicated escape route(s) or known route(s) are used exclusively for fire drills (7.5\% and 13.1\% respectively), with $p$-val$<$0.001. The issue of staff shortages was not dependent on categorical indicators. 

\begin{table}[h!]
\centering
\footnotesize
\caption{Relation between predefined issues and indicators, $p$-values of $\chi^2$ tests (significant $p$-values in bold)}
\label{tab:probchisq}
\begin{tabular}{l|c|c|c|c}
& Slow & Frightened & Shortage & Waiting for \\ 
& movement & children & of staff & free staircase \\ \hline
Organization of classes according to children’s ages (\#5)			& 0.591 & 0.092 & 0.557 &  \bf{0.027} \\
Maximum level location of spaces intended for children (\#6)			& 0.824 & 0.060 & 0.189 & \bf{$<$0.001}\\  
External escape staircase(s) available (\#7)		& 0.217 & 0.251 & 0.138 &  \bf{0.043} \\ 
Frequency of fire drills (\#11)		& 0.484 & \bf{0.015} & 0.897 & 0.232  \\ 
Announcement of fire drills to staff (\#12)				& \bf{0.023} & 0.615 & 0.691 & 0.227  \\ 
Information and instruction source for staff (\#13)			& 0.629 & 0.819 & 0.403 & 0.177 \\ 
Type of warning signal (\#14)			& 0.117 & 0.874 & 0.889 & 0.641  \\ 
Warning signal distribution (\#15)		& 0.166 & 0.661 & 0.742 & 0.602 \\ 
Exit strategy during the pre-movement phase (\#16)	& 0.722 & 0.840 & 0.558 & 0.179   \\ 
Organization during the movement phase (\#17) 	& 0.750 & 0.629 & 0.907 & 0.982  \\ 
Usage of escape routes (\#18) 		& 0.693 & 0.879 & 0.976 & \bf{$<$0.001} \\
Evaluation after fire drill completed (\#19)			 		& 0.322 & 0.846 & 0.549 & 0.239 \\  \hline
\end{tabular}
\end{table}

\section{Discussion}\label{sec:discuss}
Fire drills are currently the most widely used tool for fire training, often required by law \cite{gwynne_enhancing_2019}; nevertheless, they can be considerably limited in terms of efficiency because of various direct and indirect costs and constraints \cite{menzemer_scoping_2023, gwynne_future_2020}. In this context, early childhood educational institutions are places that, because of very specific personal, operational, and environmental conditions, show promise for maximizing the benefits of regular fire drills and to effectively perform them, unlike other types of facilities where drills may be disruptive or reduce productivity. Fire drills experienced in young age seem to have also a positive effect throughout a person's life \cite{menzemer_fire_2024}. That said, these institutions accommodate very vulnerable populations of preschool children that are potentially more sensitive to stress, risk of injury, and other issues such as adult staff's responsibilities for children's safety. Therefore, special care should be taken when fire drills are designed, executed, and evaluated in this type of environment, not least to support they are not omitted in routine practice \cite{murozaki_study_1985, taciuc_determining_2014} Recommendations for performing fire drills in early childhood educational institutions are discussed below.\par

\subsection{Frequency of fire drills}
The legislative requirements for the regular performance of fire drills in early childhood education institutions differs from country to country, but fire drills are often required annually, if not more frequently (Table~\ref{tab:freq}) \cite{page_prevalence_2015}. For example, although fire drills are not explicitly required by Czech law in nursery schools, they are generally incorporated into the fire safety policies and safety management procedures of these institutions. The results of this survey showed that 76.5\% of respondents conducted fire drills, with annual fire drills being a common frequency (82.2\% of these respondents). Based on results from this survey, more frequent fire drills (e.g., on a monthly or quarterly basis) might foster adequate emergency preparedness for both staff members and children, especially in educational institutions with frequent child or staff turnover. The literature has shown that repetition of fire training can have positive effects on improving the effectiveness of evacuation procedures and emergency responses in schools \cite{marques_key_2014, hamilton_human_2017, larusdottir_step_2011, najmanova_experimental_2023}, and our results confirmed that nursery schools that performed fire drills annually or more often encountered significantly fewer problems during the drills. Our results also indicated that children were more likely to be distressed when fire drills were performed less frequently. These findings are in line with behavioral theories that suggest that learning processes are very age-specific \cite{singer_piaget_1996, berk_child_2006} and that preschool children can reduce their fear of unknown and fire situations and learn safe practices more easily if they are exposed to such situations regularly and frequently \cite{chalmers_improving_2000}.
Therefore, we recommend that commonly required annual fire drills should ideally be complemented by additional fire drills and other fire training activities performed in a close relation to their actual purpose and the amount of information provided to the participants.  

\subsection{Purpose of fire drills and how fire drills are announced}
The value of performing fire drills can be undermined by assessing the performance of the drill and doing evacuation training simultaneously \cite{gwynne_enhancing_2019}; therefore, the purpose of a fire drill should always be clearly defined.
In addition to performing more fire drills, initial fire drills should focus especially on training with the goal of familiarizing all participants with the importance and goals of drills as well as to acquaint those involved with emergency policies and evacuation procedures specific to a school (e.g., evacuation routes in the school, meeting place/place of safety, locations, warning signal used). 
To reach the objective of training, fire drills should generally be announced to all adults involved (staff members, employees, legal guardians) and be carefully explained to children in advance. This approach can help prevent the \quotes{cry-wolf effect} \cite{rigos_cry_2019}, a phenomenon in which frequent false alarms without an actual threat can lead to poor evacuation efficiency during real emergencies. Results of our study revealed that fire drills were announced to staff in some detail at 84\% of schools, and these schools reported smoother exit evacuation processes. 
When participants have fire drill experience, the level of announcement (meaning the extent to which fire drills are announced in advance) may be gradually decreased in other drills during a school year, with the goal in a final fire drill being evacuation procedure evaluation \cite{page_prevalence_2015}, though our findings indicate that care should be taken to take each individual school's environment into consideration. 
The goals and level of announcement considered must always reflect the characteristics, needs, and possible limitations of participants involved as well as be in line with teaching attitudes of the responsible staff. In accordance with the specific context of a fire drill and the established methods and practices at each school, an attention should be also paid to communication with legal guardians of children \cite{lehna_impact_2014}.
It can be assumed that step by step training for unexpected evacuation situations along with a clear motivation and effective education can considerably help to prevent undesirable effects such as stress or anxiety by children in both training and real emergency. 

\subsection{Evacuation scenario and scheduling of fire drills}
One of the key downsides to fire drills mentioned in the literature is the possibly non-representative boundary conditions if evacuation scenarios are repeated or if they focus more on comfort/avoiding potential risks than on valuable training or evaluation processes \cite{kinateder_where_2021, menzemer_scoping_2023}. Although fire drills are always a simplification of real situations, they should ideally emphasize realistic, various conditions to reflect the diversity of actual fire events (e.g., including simulated causes and location of a fire, different availability of escape routes and safety meeting point).\footnote{Risks and fire hazards should be considered in accordance with fire safety documentation and the person responsible for fire prevention in a school.}
Existing research has demonstrated that going downstairs is the most challenging task for preschool children evacuating through buildings \cite{najmanova_experimental_2023}. This matches our results; the most frequent issue encountered by nursery schools during fire drills, according to respondents, was slow evacuation movement due to the individual assistance required by (especially younger) children. In our study, nursery schools with only ground floor spaces encountered significantly fewer issues, indicating that schools on ground floors or in one-story buildings might be considered preferable in nursery schools designing, especially if younger children (e.g., under four years of age) are involved \cite{murozaki_study_1985}.
Our findings also support stricter fire safety requirements for early childhood education institutions located in multi-story buildings or limiting building heights for these types of facilities as prescribed in some countries \cite{page_prevalence_2015}. Because the efficiency of evacuations (notably on stairs) can be greatly affected by how familiar children are with escape routes and environments \cite{murozaki_study_1985, larusdottir_evacuation_2012, najmanova_experimental_2017}, our results point towards including the use of escape routes that are not used daily and/or are unfamiliar to children (e.g., external escape staircases) in fire training and drills. \par
While it is a not generally unusual to schedule fire drills during \quotes{favorable} hours, the literature demonstrated that real evacuations occur more evenly throughout the day \cite{kinateder_where_2021}. Experimental research on the evacuation of preschool children has indicated that reactions by children and staff can differ considerably depending on the time of day, their location in a school, actual staff-to-child ratios, and the activities being carried out when a fire drill starts. Various evacuation procedures may be required because of different levels of attention and potentially limited willingness to evacuate children have; a special case is nap time \cite{najmanova_experimental_2023-1}.
With regard to relatively firm timetables in early childhood education institutions, different days of the week and times of day should be assumed for practicing and training evacuation processes. 
Seasonal variations (such as having indoor/summer clothing versus winter clothing) have also been shown to impact evacuation speeds and strategies \cite{kholshevnikov_pre-school_2009}. 
In our study, received comments indicated that evacuation training in more challenging conditions need not be preferred because of concerns about anxiety and stress to children. Hence, in addition to regular fire drills, separate training exercises focused on various actions related to building evacuation would likely be a valuable source of experience for all participants for future decisions in drills and real emergencies. 
It is also interesting to note that, based on survey comments, it is not unusual for nursery schools to organize fire drills in the form of games with special motivational rewards adjusted to be age- and child-appropriate. Such alternative approaches may have a number of advantages (such as stress reduction for children) \cite{cohen_children_2012} and gamification of fire drills and education for children should likely be encouraged \cite{capuano_knowledge-based_2015, al-smadi_decoupling_2018}. However, evaluating how this approach corresponds with actual real-life scenarios is still necessary to assess its effect on training and  its role in enhancing emergency preparedness of participants. 

\subsection{Evaluation of fire drills}
Routine fire drills can suffer from inconsistent evaluation requirements \cite{kinateder_where_2021}. Current research has also investigated thorny issues involving \quotes{successful} fire drill assessment including reliable benchmarks and evaluation indicators \cite{gwynne_enhancing_2019}. Our study revealed that a debriefing after fire drills was carried out in more than 97\% of nursery schools (in 61.1\% of these cases, in writing). However, the content and form of debriefs was not probed because there is no official, standardized report template available for Czech nursery schools. 
We can only recommend based on our results and prior research that fire drills be reviewed and evaluated in a brief and easy to complete form, to move towards fire drill report standardization \cite{baig_empirical_2024} taking specific requirements for nursery schools into consideration \cite{page_prevalence_2015}. 
In addition to baseline data (day, time, and weather), evacuation context (evacuation purpose and scenario, routes and available place of safety), equipment performance, and total evacuation time, attention should be also given to initial activities of the participants, staff-to-child ratios, and staff and (particularly young) children behaviors during a drill. Potential inconveniences and gaps in evacuation procedures should likely be properly documented to form the basis of resulting improvements. Completed fire drills should also be discussed with participating staff and children (e.g., in a form of daily activities), and reports should be made available to new staff joining a school. Subsequent communication with legal guardians of children is desirable to ensure further feedback and discussion in the home environment \cite{beckett_implementing_2014}. \par

\subsection{Fire safety education}
Fire drills that include both training and evaluation represent a part of a comprehensive approach to fire safety education. Since evacuation processes in early childhood education institutions are to a large extent dependent on the decisions and actions taken by responsible staff \cite{kholshevnikov_study_2012, najmanova_experimental_2023-1}, regular fire protection education for staff should be taken seriously.
Our study confirmed that most Czech nursery school staff are trained by qualified fire prevention professionals.\footnote{In Czechia, the personnel in early childhood educational institutions must receive regular fire education (at least every two years); staff members in charge of fire patrols must be specially trained \cite{noauthor_act_1985}.} 
Therefore, joint cooperation between fire safety experts, psychologists, and fire rescue services should lead to the development of a general framework of fire safety education that complies with the specifics of fire safety and evacuation of early childhood education institutions. In addition, combination with other fire training methods may bring about advantages \cite{menzemer_scoping_2023, gwynne_enhancing_2019}. \par
Although preschool safety is largely in the hands of caregivers \cite{page_prevalence_2015}, the importance of fire safety education for children, tailored to incorporated age specifics, should not be underestimated. Children with different ages need different teaching styles \cite{cote_fire_2002, borzekowski_young_2013}. Preschool-aged children may be limited in reasoning, manipulating information, and relating events together \cite{siegler_childrens_1998, shaffer_developmental_2010}. Therefore, according to our findings and the literature, appropriate and effective methods for explaining evacuation procedures and fire drill objectives to children should be carefully adapted to their needs and age in education programs \cite{mcconnell_evaluation_1996, secer_fire_2006}, for example, by employing various resources and techniques (so-called multi-messages/multiple methods approach) \cite{lehna_impact_2014}. 
Involving fire escape planning and drills in child fire safety education and gamification can considerably reduce stress and streamline evacuation processes \cite{pooley_evidence-based_2021}. Since understanding evacuation procedures is a key factor affecting the speed of evacuation \cite{murozaki_study_1985}, children should be familiarized with the main principles of emergency evacuation, be aware of the type of warning signal and its meaning, the type of instructions that they can subsequently expect, and the location of the safety meeting point. The opportunity to get acquainted with the evacuation routes and signs that are not used daily and present in a school during pre-instructions and other educational activities can be important to help children understand safety procedures in a systematic and gradual manner and increase their positive motivation for regular evacuation training.

\section{Conclusions}

This article presented data from an anonymous online survey focused on the performance of fire drills in nursery schools in Czechia for children 3~to 6~years of age. Respondents from \numprint{1151} nursery schools (23.5\% of nursery schools invited, 21.5\% of all officially registered nursery schools in Czechia) took part in the survey. The survey provided insight into the operational conditions, experiences with safety procedures, and conduct of fire drills in nursery schools, enabling us to provide informed suggestions for improving the effectiveness of fire drills in preschools to various stakeholders. Results confirmed that nursery schools in Czechia represent facilities with very different and unique environmental conditions (e.g., location/building, capacity, and other occupancy characteristics).
Our survey findings showed that nursery schools implementing fire drills tended to have a significantly larger number of classes, a lower number of staff per class, and a higher ratio of staff to children compared to those that did not conduct such drills. Nursery schools that conducted fire drills less than once a year faced notably more problems than those that held them at least once a year or more frequently. Furthermore, our research indicated that nursery schools located only on the ground floor experienced fewer problems during fire drills compared to schools with rooms on the first floor or higher. The most frequent evacuation issue encountered during fire drills reported by respondents was slow movement during evacuation because many children needed help, especially on stairs.

Our study supported prior research and suggested that fire drills in preschools can be an efficient tool for emergency preparedness if specific aspects of their design, execution, and evaluation are taken into account and if the characteristics, needs, and possible limitations of children are carefully incorporated into drills. 
While limited to the Czech context, this study confirmed many findings from prior studies and provided a valuable background for a wider international application in fire safety practices and building design. 
This study also adds richness to the corpus of research related to children evacuation, particularly emphasizing the need to consider age-specific factors that can impact evacuation efficiency and provides a foundation for applying these insights beyond fire-related evacuation scenarios.

However, our research findings must be interpreted in light of several existing limitations, including challenges in data collection, potential bias in sample size, non-response and response bias, and cultural and organizational factors related to the Czech setting that may affect generalization of results. Furthermore, we recognize a limitation in the survey's scope stemming from a necessary trade-off between obtaining desired data and reducing survey fatigue for participants. Future investigations on the specifics in the structural, functional, and accessibility features of schools (including emergency exits, construction types, daily operations, and facilities for children with disabilities or special needs) can provide additional context relevant to the guiding of fire safety strategies tailored to nursery schools. In addition, understanding more details regarding the execution of fire drills (such as timing, participant reactions, emotions, and perceptions) would enhance the development of fire drill design and broaden our comprehension of the environmental and operational factors that affect fire safety in nursery schools.

\bibliography{Drills}

\end{document}